# Topological boundary states in engineered quantum-dot molecules on the InAs(111)A surface

Van Dong Pham[1], Yi Pan[1,2], Steven C. Erwin[3], Felix von Oppen[4], Kiyoshi Kanisawa[5], and Stefan Fölsch[1,*]

*[1] Paul-Drude-Institut für Festkörperelektronik, Hausvogteiplatz 5-7, Leibniz-Institut im Forschungsverbund Berlin e. V., 10117 Berlin, Germany*

*[2] Center for Spintronics and Quantum Systems, State Key Laboratory for Mechanical Behavior of Materials, Xi'an Jiaotong University, Xi'an 710049, China*

*[3] Center for Computational Materials Science, Naval Research Laboratory, Washington, DC 20375, USA*

*[4] Dahlem Center for Complex Quantum Systems and Fachbereich Physik, Freie Universität Berlin, 14195 Berlin, Germany*

*[5] NTT Basic Research Laboratories, NTT Corporation, 3-1 Morinosato Wakamiya, Atsugi, Kanagawa, 243-0198, Japan*

## Abstract

Atom manipulation by scanning tunneling microscopy was used to construct quantum dots on the InAs(111)A surface. Each dot comprised six ionized indium adatoms. The positively charged adatoms create a confining potential acting on surface-state electrons, leading to the emergence of a bound state associated with the dot. By lining up the dots into $N$-dot chains with alternating tunnel coupling between them, quantum-dot molecules were constructed that revealed electronic boundary states as predicted by the Su-Schrieffer-Heeger (SSH) model of one-dimensional topological phases. Dot chains with odd $N$ were constructed such that they host a single end or domain-wall state, allowing one to probe the localization of the boundary state on a given sublattice by scanning tunneling spectroscopy. We found probability density also on the forbidden sublattice together with an asymmetric energy spectrum of the chain-confined states. This deviation from the SSH model arises because the dots are charged and create a variation in onsite potential along the chain – which does not remove the boundary states but shifts their energy away from the midgap position. Our results demonstrate that topological boundary states can be created in quantum-dot arrays engineered with atomic-scale precision.

## Orcid IDs

V.D.P.: 0000-0002-1416-2575;  Y.P.:  0000-0003-1978-475X;  S.C.E.: 0000-0002-9675-9411; F.v.O.: 0000-0002-2537-7256; SF: 0000-0002-3336-2644

---

*foelsch@pdi-berlin.de

# I. INTRODUCTION

Topological insulators [1,2] are a class of materials defined by a remarkable feature: electronic states at their surfaces are protected by topology and thus cannot be eliminated without closing the energy gap of the insulator. Most research has focused on naturally occurring topological insulating materials [3,4,5,6,7]. But topological states can also be created artificially in the laboratory, by applying design principles derived from the fundamental theory that governs their unusual nature. An example of such a theory is the Su-Schrieffer-Heeger (SSH) model [8,9,10]: it describes a chain of degenerate orbitals in which electrons hop with alternating amplitude between nearest neighbors. This model captures the properties of end and domain-wall states in a dimerized chain [11], exemplifying a prototypical case of protected in-gap states at the boundary between distinct topological phases in one dimension. Different experimental approaches were used to realize the SSH model ranging from classical systems [12,13,14] to ultracold quantum gases [15,16], and photonic metamaterials [17,18]. In addition, engineered nanoscale materials were utilized such as graphene nanoribbons [19,20] and polymer chains [21] created by on-surface synthesis, self-assembled atomic chains on vicinal Si surfaces [22], as well as chains of surface vacancies [23,24] and subsurface dopants [25] generated by scanning-probe techniques. The general goal of these efforts is to tune the boundary states by proper design of the materials.

In our previous work [26], we used the probe tip of a scanning tunneling microscope (STM) to create quantum dots on the InAs(111)A surface each consisting of a short, six-atom chain of ionized adatoms. The positive charge of the native adatoms led to the confinement of surface-state electrons [27] and the emergence of a bound state associated with the dot. By lining up these dots ("artificial atoms") into $N$-dot chains and creating alternating long and short spacings between them we constructed quantum-dot molecules that revealed electronic boundary states as they are predicted by the SSH model of finite dimer chains.

In this paper, we focus on the same physical system and address the specific case of dimerized chains consisting of an odd number $N$ of dots. This situation provides new insights because it allows one to realize chains hosting only a single boundary (end or domain-wall) state. In the SSH model, a given boundary state is expected to reside only on one sublattice which can be readily tested experimentally by scanning tunneling spectroscopy (STS). Secondly, the molecular-state energies observed for dimerized chains with odd $N$ can be directly correlated with the band structure of an infinite dimer chain, revealing that the observed end states are indeed in-gap states. Deviations from the SSH model will be discussed regarding the sublattice localization of the end states and their actual energy position within the gap. The deviations arise because the dots are charged and create a variation in onsite potential along the chain − which is a defining aspect of the present physical system.

## II. EXPERIMENTAL METHODS

An ultrahigh vacuum (UHV) STM system operated at 5 K was used to carry out experiments on InAs(111)A grown by molecular beam epitaxy (for details on sample growth and preparation see Ref. [27]). STS measurements of the differential tunneling conductance $dI/dV$ were performed at constant tip-surface separation, using a lock-in amplifier at a peak-to-peak modulation voltage of 5 mV (unless stated otherwise) and a modulation frequency of 675 Hz. Spectra in this work show the normalized conductance $(dI/dV)/(I/V)$ providing an approximate measure of the surface density of states [28]. We used electrochemically etched tungsten tips cleaned by Ne ion sputtering and electron beam heating. A final conditioning of the tip was accomplished by *in situ* voltage pulsing [29] to agglomerate indium at the tip apex. This tip state makes it possible to exchange individual In atoms between the surface and the scanning-probe tip with a high degree of reliability [30].

## III. RESULTS AND DISCUSSION

The left hand-side STM image in Fig. 1(a) shows a quantum dot consisting of six In adatoms residing on nearest-neighbor vacancy sites of the (2×2)-reconstructed surface at an interatomic spacing of $a'=\sqrt{2}a_0=8.57$ Å, with $a_0=6.06$ Å the cubic lattice constant of InAs. The increased apparent height around the dot is due to the screened Coulomb potential arising from the positive charge associated with the adatoms. The local potential confines surface states of pristine InAs(111)A, giving rise to a bound state with a single-lobed probability density at ~0.1 eV below the Fermi level (at sample bias $V$=0) [27]. In the green spectrum in the upper panel of Fig. 1(b), the bound state is manifested by the discrete conductance peak at -90 mV. This state will be utilized in the following to realize artificial molecular states in quantum-dot chains of precisely defined configuration.

 The right hand-side STM image in Fig. 1(a) shows two identical $In_6$ dots placed side by side at a spacing of 6 in units of $\sqrt{3}a'$ along the $\langle 211 \rangle$ in-plane direction. The electronic coupling between the dots leads to a bonding ($\sigma$) and an antibonding state ($\sigma*$) and thereby two conductance peaks in the corresponding red spectrum, in analogy to our earlier findings for dimers of In adatom chains on InAs(111)A in side-by-side and other arrangements [26,27,30]. For the specific spacing chosen here, the $\sigma-\sigma*$ splitting amounts to $\Delta_{\sigma-\sigma*}$=45 mV. The spectra also show that the $\sigma-\sigma*$ doublet (red curve) is downshifted from the energy of the single dot (green curve) by a value of $\Delta_s$=−16 mV [31] at the interdot spacing of 6. We interpret this observation as an electrostatic shift arising from the potential change that each dot experiences from the other [26]. The $\sigma-\sigma*$ splitting is analyzed in closer detail in Fig. 1(b) which adds contour plots of conductance spectra recorded at an interdot spacing of 3 to 10 units. Consistent with their symmetric ($\sigma$) and antisymmetric ($\sigma*$) wave-function character, both the $\sigma$ and $\sigma*$ states are observed when the tip position is fixed on the dots (center panel) while only the $\sigma$ state is detected when placing the tip precisely between the dots (lower panel). As expected, the electronic coupling between the dots (and thereby the $\sigma-\sigma*$ splitting) decreases as the interdot spacing increases.

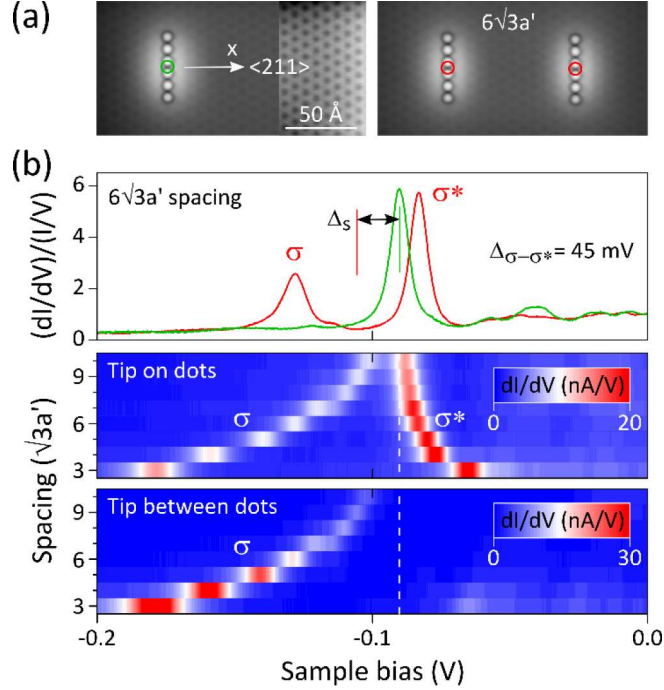

Fig. 1. (a) Left panel: STM topography image (0.1 V, 0.1 nA) of a chain of six In adatoms placed on adjacent vacancy sites of the (2×2)-reconstructed surface. The chain behaves like a quantum dot because of the positive charge of the adatoms. The enhanced contrast on the right highlights the surface vacancies. Right panel: same area after adding an identical dot at a spacing of 6 √3a′ along the ⟨211⟩ in-plane direction with a ′=8.57 Å the surface-lattice constant. (b) Top panel: normalized conductance spectra recorded with the STM tip probing the discrete dot (green curve) and the dots of the dimer (red curve); a single bound state is found for the discrete dot while bonding (σ) and antibonding states (σ∗) emerge for the dimer. Δ_s denotes the electrostatic shift of the σ−σ∗ doublet with respect to the bound-state peak. Center and lower panel: contour plots of the conductance as a function of sample bias (x axis) and dot spacing (y axis) varied between 3 and 10 in units of √3a′ recorded with the tip placed above the dots (center panel) and in between of the dots (lower panel); the vertical dashed line marks the peak position of the bound state.

In this work, we investigate dimerized chains consisting of an odd number of $In_6$ dots. Figure 2(a) shows a dimerized chain of nine dots with alternating interdot spacing of 4 and 3. For a simple tight-binding treatment with adjacent hopping, we define hopping amplitudes $t=e\Delta_{\sigma-\sigma*}/2$ based on the σ−σ∗ splittings observed in dimers [see Fig. 1(b)], explicitly $t(3)=55$ meV and $t(4)=39$ meV for the spacings of 3 and 4 units in Fig. 2, respectively. To describe this chain with the SSH model [11], we also define a unit cell as highlighted by the yellow box in panel (a) which contains the leftmost dimer of the chain. Here, the intracell hopping is smaller than the intercell hopping, $t(4)<t(3)$. This implies the existence of an end state with a probability density that is maximal on the leftmost site (here belonging to sublattice $A$) and exponentially decaying along the sites of the same sublattice [11]. On the other hand, the rightmost unit cell highlighted by the dashed-line box has a vacancy (no dot) at site $i$=10. For a finite SSH chain with five unit cells and weakly coupled end sites this situation can be modeled by steadily increasing the onsite energy of *one* end site while leaving it zero for *all other* lattice sites (Appendix A, Fig. 7): in doing so, the odd superposition of the two end states is first pushed out of the energy gap and eventually out of the entire level spectrum at large onsite energy, leaving only one end state at in-gap energies.

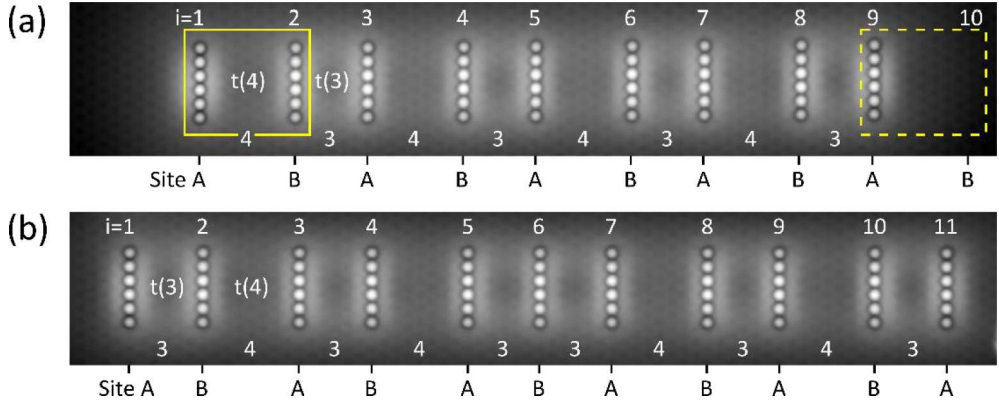

*Fig. 2. (a) Topography image (0.1 V, 0.1 nA) of a dimerized chain of nine dots with alternating spacing of 4 and 3 in units of √3a'. The solid-line box highlights the unit cell at the left hand-side end of the chain containing one site of sublattice A and one site of sublattice B; the unit cell at the right hand-side end (dashed-line box) has a vacancy (no dot) at site index i=10. The resulting configuration has a weakly (strongly) coupled dot on the left (right). (b) Topography image (0.1 V, 0.1 nA) of a dimerized chain of 11 dots with alternating spacing of 4 and 3 units plus an internal domain wall at site i=6; the resulting chain has strongly coupled dots at both ends and a heavy domain wall with strong coupling between the dot at site i=6 and its neighbors at i=5 and 7. t(3) and t(4) in panels (a) and (b) denote the hopping amplitudes $t=e\Delta_{\sigma-\sigma*}/2$ with $\Delta_{\sigma-\sigma*}$ the $\sigma-\sigma*$ splitting observed in discrete dimers at a dot spacing of 3 and 4 units, respectively [see Fig. 1(b)].*

The end of this chain is actually just a particular case of a domain wall. A domain wall can also be introduced somewhere along the chain [23,24]. As an example, Fig. 2(b) shows a chain of 11 dots with an internal domain wall at site *i*=6 constituting a boundary between the two possible dimerizations. For this chain the SSH model predicts there are no midgap states associated with the ends, because the outermost dots at both ends are strongly coupled to their nearest neighbor. On the other hand, the domain wall does come with a state resembling the first excited state of an isolated trimer [11] for the "heavy" domain-wall configuration in Fig. 2(b).

The odd-*N* chains investigated in this work were designed to host a single end state (or domain-wall state) as shown by the examples in Fig. 2. The general features of these boundary states can be directly verified by STS measurements. Starting with the dimerized chain of nine dots in Fig. 2(a), we recorded the differential tunneling conductance with the tip scanning at constant height and fixed sample bias along the chain axis as marked by the dashed line in the topmost STM image in Fig. 3(a). Performing this scan as a function of bias yields the conductance map $D(x,V)$ depicted below the STM image revealing that the level spectrum of the chain-confined states spans a range of ~200 mV, roughly from -70 to -270 mV. The end state denoted *e* associated with the left end site of the chain is highlighted by the white box in the middle of the level spectrum; the corresponding spatial conductance map $D(x,y)$ is shown at the bottom of Fig. 3(a). In the SSH model, the midgap state lives only on one sublattice (here the *A* sites) and decays into the interior of the chain. The experimental data show that this is approximately but not exactly true for the real structure studied here because a small but nonvanishing conductance (probability density) is detected also on the *B* sites; this discrepancy will be further discussed below in context with Fig. 5.

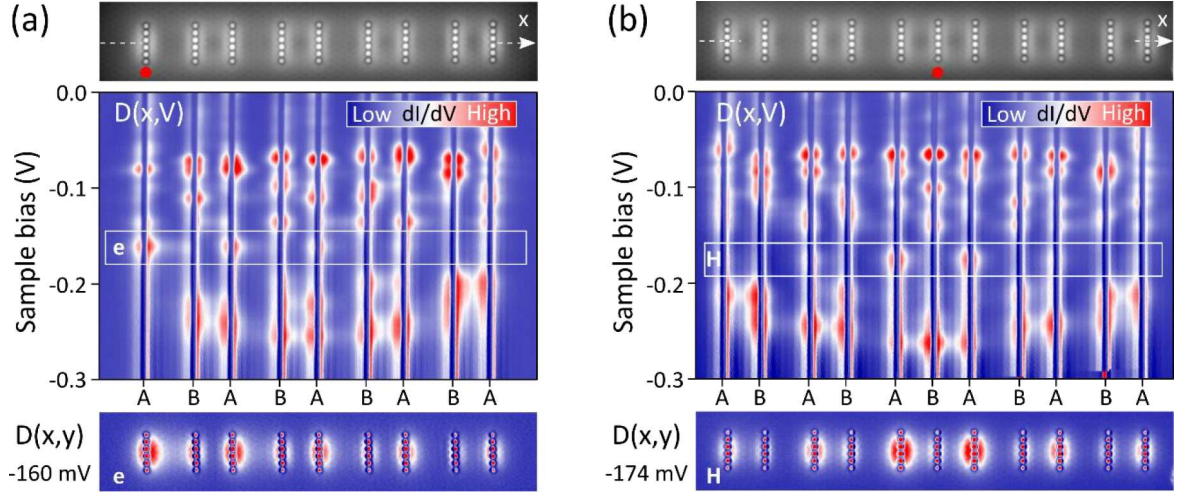

*Fig. 3. (a) Top panel: topography image (0.1 V, 0.1 nA) of the same nine-dot chain as in Fig. 2(a). The red marker indicates the weakly coupled end site on the left. Center panel: conductance map D(x,V) showing the dI/dV signal recorded at constant tip height along the symmetry axis of the dimerized chain (dashed line denoted x in the STM image) and as a function of sample bias V. Chain-confined states occur over a bias range of roughly -70 to -270 mV. The boundary state (denoted e) associated with the weakly coupled end site emerges in the middle of the level spectrum as highlighted by the white box. Lower panel: spatial conductance map D(x,y) of the boundary state recorded at -160 mV revealing that the probability density is maximum at the end site on the left and decays along the A sites towards the right. (b) Same information as in panel (a) but for the 11-dot chain as shown in Fig. 2(b). The boundary state denoted H is associated with the heavy domain wall at site i=6 and again occurs in the middle of the level spectrum, see the D(x,V) map in the center panel. The density distribution of the boundary state resembles that of the first excited state of an isolated trimer, see the spatial D(x,y) map in the lower panel. All conductance maps were recorded at a lock-in modulation voltage of 10 mV (peak-to-peak).*

Similar data is presented in Fig. 3(b) for the 11-dot chain with an internal domain wall as shown in Fig. 2(b). It is evident that the boundary state associated with the domain wall (i) occurs roughly in the middle of the observed energy level spectrum, and (ii) has maximum probability density on the two dots adjacent to the center dot (domain-wall position) at site $i$=6. This resembles the first excited state of an isolated linear trimer $\psi_2$=($-1/\sqrt{2}$, 0, $+1/\sqrt{2}$) in the basis of the three dots, as predicted by the SSH model in the fully dimerized limit [11]. Aside from this heavy domain-wall case, also a light domain wall can be constructed which then gives rise to a boundary state localized exclusively on the center dot [32].

For closer inspection of the confined-state energies, Fig. 4 displays spatially averaged conductance spectra of dimerized chains with $N$=5, 7, and 9 dots. The left panel shows spectra of chains with an alternating spacing of 4 and 3 units, like the $N$=9 chain in Fig. 2(a). For convenience, chains of this kind are termed (4,3) chains in the following. The right panel shows equivalent spectra of (5,3) chains having an alternating spacing of 5 and 3 units. [The spacing of 5 units implies a hopping amplitude of $t(5)$=29 meV as deduced from the corresponding $\Delta_{\sigma-\sigma*}$ splitting in dimers, see Fig. 1(b).] The data of both configurations reveal spectral features similar to those we observed earlier for quantum-dot-dimer chains on InAs(111)A [26]: a series of conductance peaks reflecting the molecular states of the chain together with an extra peak denoted $P$ revealing a state at higher energy. It appears likely that this pronounced spectral peak derives from accumulated electrons near the InAs surface [33,34,35,36] that become laterally confined in the presence of the quantum-dot chain. This state lies ~70 meV below the Fermi level, largely insensitive to the number of dots and their detailed configuration. In contrast, the level spectrum of the molecular states evolves as $N$ is increased, undergoing a uniform downward shift in energy. The end state $e$ (marked red) occurs roughly in the middle of the gap. The $(N-1)/2$ molecular levels below the end state are clearly seen in the spectra − in particular for the (4,3) configuration − whereas the conductance peaks of the topmost levels above the end state merge with the extra peak $P$, making it impossible to capture these levels experimentally.

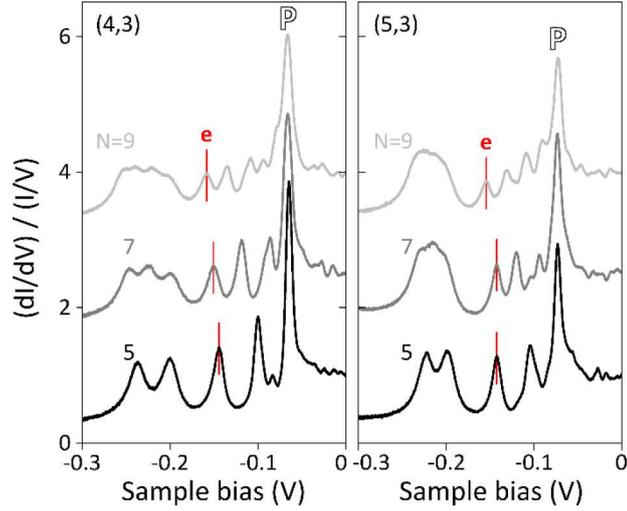

Fig. 4. Left panel: normalized conductance spectra (dI/dV) / (I/V) for dimerized chains with odd number N of dots and alternating spacing of 4 and 3 units, termed (4,3) configuration; the spectra were spatially averaged over all dots of the chain. The boundary state e associated with the weakly coupled end site is marked by a vertical red bar. A total of (N−1)/2 molecular levels below the boundary state is resolved in the spectra whereas the conductance peaks of the levels above the boundary state merge with the spectral feature at higher energy (denoted P), presumably originating from accumulated electrons near the InAs surface. Right panel: same spectral information for dimerized chains with odd N and in (5,3) configuration. As observed for the (4,3) chains, the level spectrum of the molecular states successively evolves as N is increased while the state associated with the extra peak P is largely insensitive to the number of dots.

Further information can be extracted from site-resolved conductance spectra obtained by probing specific dots of the dimer chain. Two exemplary chains in (4,3) and (5,3) configuration consisting of $N$=7 dots are illustrated in Fig. 5(a). The respective $A$ and $B$ sites are assigned to colors as drawn in the STM images and the corresponding spectra are shown in Fig. 5(b): for the (4,3) chain (left panel), six molecular levels with quantum numbers $n$=1 to 6 are well resolved while the topmost level ($n$=7) is obscured by the extra peak. Of special interest is the end state ($n$=4) associated with the weakly coupled end site on the left of the chain belonging to sublattice $A$. The observed probability density is maximum on this end site and decays along the adjacent $A$ sites of the dimer chain, in qualitative agreement with the SSH model. Nevertheless, there is also nonvanishing probability density on the $B$ sites, as already noted in the discussion of Fig. 3(a). In addition, the observed level spectrum is not symmetric about its center. Both these effects appear to be even more pronounced for the (5,3) chain made of seven dots as evident from the related conductance spectra displayed in the right panel of Fig. 5(b).

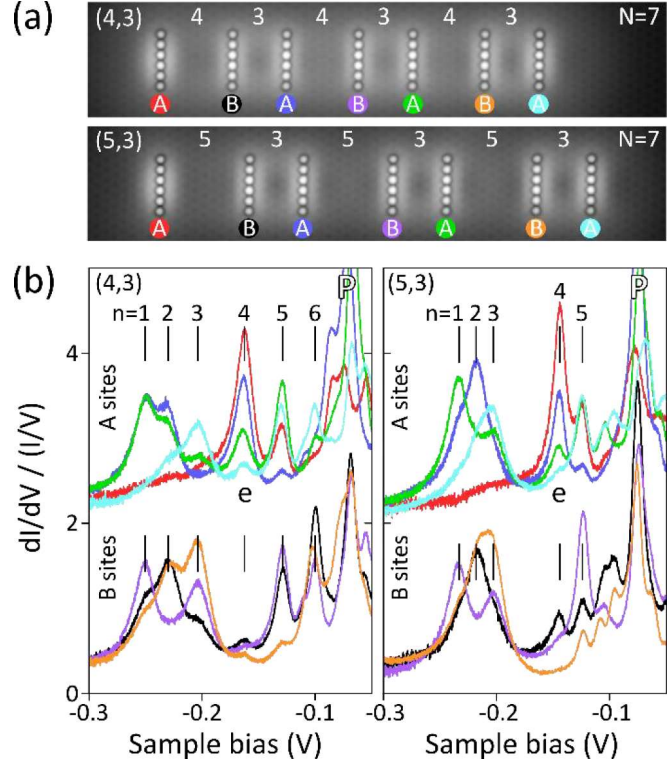

*Fig. 5. (a) Topography images (0.1 V, 0.1 nA) of dimer chains with N=7 dots in (4,3) [top] and (5,3) configuration (bottom) having a weakly coupled dot at the left end, the respective A and B sites probed by the STM tip are indicated by colors. (b) Corresponding spectra for (4,3) [left] and (5,3) configuration (right); vertical bars together with quantum numbers n mark those molecular states which are unambiguously identified. In both cases, the boundary state e (n=4) associated with the left end of the chains shows conductance (probability density) predominantly on the A sites and to a smaller extent also on the "forbidden" B sites. This deviation from the SSH model is due to the variation in onsite potential along the chain (see main text for details).*

The experimental observations in Figs. 4 and 5 reveal deviations from the properties of finite ideal SSH chains: the presence of end-state density also on the forbidden sublattice and an asymmetric level spectrum that undergoes a uniform downward shift in energy as the number of sites (dots) is increased. (Data showing the observed molecular-state energies versus $N$ is in Appendix B, Fig. 8.) These deviations arise from the positive charge associated with the dots, which creates a varying onsite potential along the chain that is higher at the ends than in the interior, breaking the sublattice symmetry. The variation in onsite potential and its effect on the level spectrum can be modeled as discussed in Appendix B and reported earlier for dimer chains of even $N$ [26]. Moreover, the model also reproduces nonvanishing end-state density on the forbidden sublattice.

The symmetry breaking induced by the variation in onsite potential does not remove the boundary states in the present case but leads to a shift in their energies. To see this explicitly for the case of end states, we apply the textbook case of a one-dimension atomic chain in which electrons hop between nearest-neighbor sites. This treatment allows one to correlate the molecular-state energies observed for dimerized chains of odd $N$ with the band structure of a dimer chain with periodic boundary conditions.

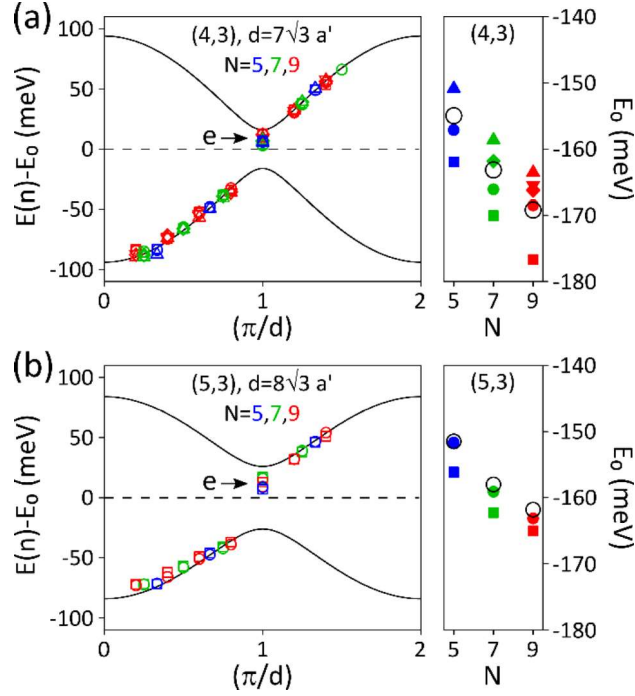

Fig. 6. (a) Left diagram: tight-binding (TB) energy bands of an infinite dimer chain (full black lines) assuming the hopping amplitudes t(3)=55 meV and t(4)=39 meV based on the σ–σ* splittings in dimers with 3 and 4 units spacing; d=7 √3 a' is the unit-cell length. Colored symbols show the energy values $E_n$ of the molecular states observed for twelve individual (4,3) chains with N=5 (green), seven (blue), and nine dots (red) plus a constant energy offset $E_0$ at given N. The energy values are plotted versus the discrete wave vector $k_n$=2πn/(N+1)d and fall on the two TB bands except the end state with n=(N+1)/2 which is in the band gap and shifted upwards relative to midgap energy. Right diagram: energy offset $E_0$ versus N (full symbols) as obtained by fitting the experimental energies to the TB bands; $E_0$ quantifies a mean electrostatic shift which can be estimated (large open circles) using a simple model that takes into account the onsite potentials along the chain (see main text for details). (b) Same as (a) but for (5,3) chains with the corresponding hopping amplitudes t(3)=55 meV and t(5)=29 meV. The end state is again in the energy gap and shifted upwards relative to the midgap position. Compared to the (4,3) case in (a), the total band width of the level spectrum is reduced and the gap size increased, in agreement with the TB model.

Consider a uniform chain with N identical atoms at constant separation a, with hopping t and onsite energy zero. In this case, the energies are $E_n = -2t \cos(\pi n / N + 1)$ with n=1,2,…,N. Each energy $E_n$ is associated with a discrete wave vector $k_n$= πn/(N+1)a [37] and hence the points $(k_n, E_n)$ fall on the continuous energy band $E(k) = -2t \cos(ka)$ of an infinite chain with interatomic separation a. When the chain dimerizes, a gap forms separating a lower band of bonding character and an upper band of antibonding character. For a finite dimerized chain of odd N, this implies that the states with n=1,…,(N–1)/2 lie on the lower band and those with n=(N+3)/2,…,N on the upper band. The remaining state with n=(N+1)/2 originates from the lone end orbital and is centered in the energy gap between the two bands.

Following this scheme, the experimental $E_n$ values can be readily compared to the tight-binding band structure of an infinite dimer chain $E(k) = \pm\sqrt{t_1^2 + t_2^2 + 2t_1 t_2 \cos kd}$ [10] with d=2a, the unit cell length. The left diagram in Fig. 5(a) shows the energy bands drawn as full black lines with the hopping amplitudes $t_1$ and $t_2$ set to the experimental values t(4)=39 meV and t(3)=55 meV corresponding to the (4,3) configuration. We fitted the experimental $E_n$ values to these bands by including all observed states except the end state with $n=(N+1)/2$ because this state is expected to be located in the gap. The only fitting parameter in the procedure is a constant energy offset $E_0$ for each chain of given $N$. $E_0$ quantifies a mean electrostatic shift resulting from the overall potential induced by the dots along the chain; this quantity can be estimated using a simple electrostatic model as outlined in Appendix C. The result of the fit is shown in the left diagram of Fig. 6(a) for twelve individual (4,3) chains consisting of five (blue), seven (green), and nine dots (red symbols), respectively. It is evident that the experimental data closely follow the tight-binding band structure. The agreement is plausible because the states associated with the bands have only moderate wave-function amplitude on the end site. In contrast, the remaining end state $e$ is found to be located in the gap and shifted upwards in energy relative to the midgap position. This is because it has maximum wave-function amplitude on the end site of weaker bonding and is thus most sensitive to the increased onsite potential at the ends. The actual $E_0$ values obtained from fitting the experimental energies are plotted as filled symbols in the right hand-side diagram of Fig. 6(a), showing fair agreement with the estimated value (empty circles). In analogy to our previous observations [26], the scatter in the experimental data (here exemplified by the filled symbols) reflects electrostatic potential disorder due to residual charged defects in the InAs(111)A substrate [38,39,40].

Figure 6(b) shows the same analysis of the band structure (left) and electrostatic shift (right) for (5,3) chains of odd $N$. In this case too, the experimental energies are in reasonable agreement with the energy bands of an infinite dimer chain, now with hoppings $t_1 \equiv t(5)$ and $t_2 \equiv t(3)$ equal to 29 and 55 meV, respectively. Likewise, the end state is in the band gap and shifted upwards relative to the midgap position. The data are consistent with a reduced total band width $2(t_2+t_1)$ and an increased energy gap $2(t_2-t_1)$ as compared to the case of (4,3) chains in Fig. 5(a).

## IV. CONCLUSIONS

We created dimerized chains of identical quantum dots on a semiconductor surface and found that these chains give rise to electronic boundary states in qualitative agreement with the SSH model of one-dimensional topological phases. Chains comprising an odd number of dots can be constructed such that they host only a single end or domain-wall state, allowing one to directly probe the sublattice character of the boundary state. Our STS data reveal nonvanishing probability density also on the forbidden sublattice. Moreover, the energy level spectrum of the manifold of molecular states observed is asymmetric about its center. These deviations from the SSH model arise because the charge associated with the dots creates a varying onsite potential that is higher at the ends than in the bulk of the chain, breaking the sublattice symmetry. The symmetry breaking does not remove the boundary states but leads to a shift in their energies. This is especially obvious for end states located in the energy band but shifted upwards from the midgap position by the increased onsite potential at the end of the chain.

In conclusion, STM-based atom-by-atom assembly in combination with scanning tunneling spectroscopy provides insight into and control of the electronic states in engineered nanomaterials at the smallest size scales. By applying the design principles derived from the SSH model, we have shown that the predicted boundary states can be emulated in quantum-dot-dimer chains on the InAs(111)A surface. At the same time, our results highlight the important role of electrostatics in these artificial quantum structures on semiconductor platform.

**ACKNOWLEDGEMENTS**

V.D.P., Y.P., and S.F. acknowledge funding by the Deutsche Forschungsgemeinschaft (DFG, German Research Foundation) under Grants No. FO362/4-2 and No. 437494632. Y.P. also acknowledges the National Key Research and Development Program of China (Grant No. 2022YFA1204100) for supporting the collaboration. S.C.E. was supported by the Office of Naval Research through the Naval Research Laboratory's Basic Research Program. F.v.O. was supported by CRC 183 of the Deutsche Forschungsgemeinschaft.

## APPENDIX A: FINITE DIMERIZED CHAINS

We consider a single-particle electronic tight-binding Hamiltonian as used in the SSH model [11] in which electrons hop with alternating amplitudes $t_1$ and $t_2$ between nearest neighbors and vanishing onsite energy $\alpha$ on each site. A finite dimer chain with five unit cells ($N$=10 sites) is shown in Fig. 7(a): each cell contains one $A$ site and one $B$ site, respectively. Topological end states emerge at midgap energy when the hopping is smaller within cells than between cells, $t_1 < t_2$. As an example, the left diagram of Fig. 7(b) shows the resulting energy level spectrum for $t_1/t_2$=0.7. Because of the finite length of the chain, the two end states overlap and form even and odd superpositions yielding the two levels $e_1$ and $e_2$ in the gap region shaded yellow. The center diagram illustrates the effect on the level spectrum as the onsite energy of the rightmost site at $i$=10 is turned on and continuously increased while keeping $\alpha$ equal to zero for all other sites: both $e_1$ and $e_2$ are raised in energy, $e_1$ converges towards the midgap position while $e_2$ is pushed out of the gap and into the upper half of the level spectrum. At even larger site energy the rightmost site eventually becomes completely decoupled from all other sites. The nine lower-lying states are then essentially equivalent to the levels of a dimerized chain consisting of $N$=9 identical sites exhibiting a single end state $e$ located at midgap energy, as shown in the right hand-side diagram of Fig. 7(b). Panel (c) adds the wave-function coefficients $c_{n,i}$ of the superpositions $e_1$ ($n$=5) and $e_2$ ($n$=6) for the starting situation with $N$=10 in (b), together with the same information for the single end state $e$ ($n$=5) for the case $N$=9. Each of these states is exponentially localized on a weakly coupled end site and resides on one sublattice.

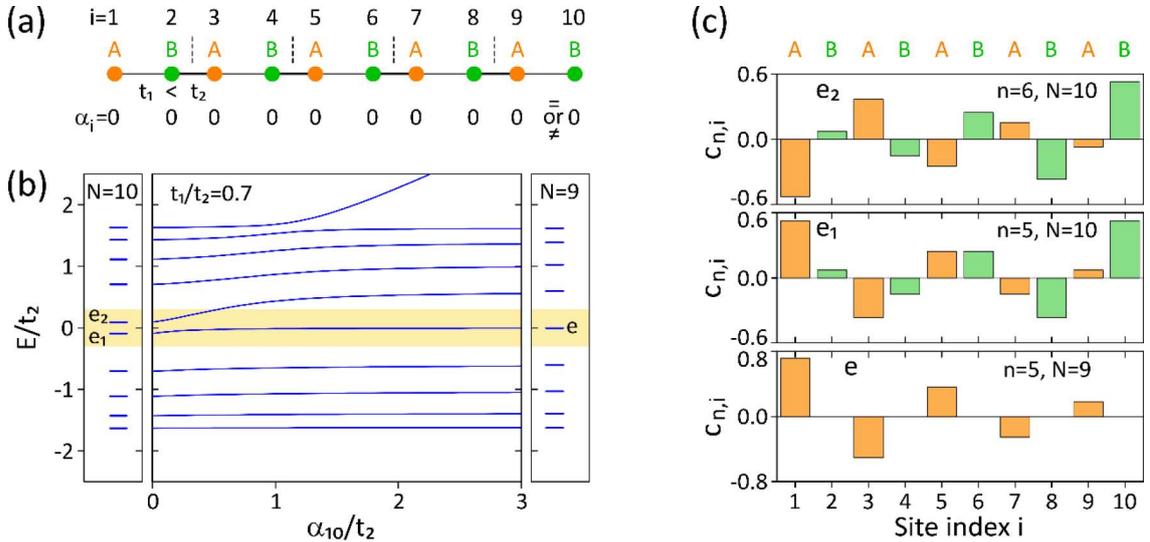

Fig. 7. (a) Sketch of a dimer chain with N/2=5 unit cells each with one A site and one B site, the hopping within cells is weaker than between cells, $t_1 < t_2$; lattice sites are numbered by index i from left to right. (b) Left panel: corresponding energy spectrum with $t_1/t_2$=0.7 and all onsite energies $\alpha_i$=0, the end states form even and odd superpositions leading to two levels $e_1$ and $e_2$ in the gap (shaded yellow) close to zero energy. Center panel: evolution of the spectrum as the onsite energy $\alpha_{10}$ of the end site on the right (i=10) is steadily increased while keeping $\alpha$ zero otherwise. Right panel: energy spectrum of a dimerized chain with N=9 sites and all $\alpha$ zero revealing a single end state in the gap at zero energy, the spectrum is equivalent to the limiting case at large $\alpha_{10}/t_2$ in the center diagram. (c) Wave-function coefficients of the superpositions $e_1$ and $e_2$ for N=10 (top and center) as well as the single end state e for N=9 (bottom).

## APPENDIX B: ENERGY LEVEL SPECTRUM OF THE DIMERIZED CHAINS

The conductance spectra in Figs. 4 and 5 revealed that the measured energies $E_n$ indicate a level spectrum that is *not* symmetric about its center. A more extensive data set is plotted in Fig. 8 showing the experimental energies (small filled symbols) observed for various individual chains with odd $N$=3,5,7, and 9 in (4,3) and (5,3) configuration, respectively. As evident, the level spectrum is asymmetric and experiences a uniform downward shift as $N$ is increased. The large empty circles display the theoretical energies obtained from a nearest-neighbor tight-binding Hamiltonian with experimental hoppings $t_1,t_2$=39 and 55 meV for (4,3) as well as $t_1,t_2$=29 and 55 meV for (5,3) configuration. The effect of electrostatics was taken into account by allowing for nonvanishing onsite energies $e\langle V_i \rangle$ defined by the expectation value $\langle V_i \rangle = \langle \varphi_i | V | \varphi_i \rangle$ corresponding to the onsite potentials. $\varphi$ is assumed to be a hydrogenic orbital located on each dot center and $V$ denotes the screened Coulomb potential from all the charged In adatoms calculated based on the result derived in Appendix B of Ref. [41]. This procedure is in analogy to our previous treatment of quantum-dot dimer chains with even $N$ [26], including the assumption that each In adatom is positively charged and incompletely ionized, $q$=0.69. Good agreement between theory and experiment is obtained also for dimerized chains of odd $N$ as illustrated in Fig. 8(a).

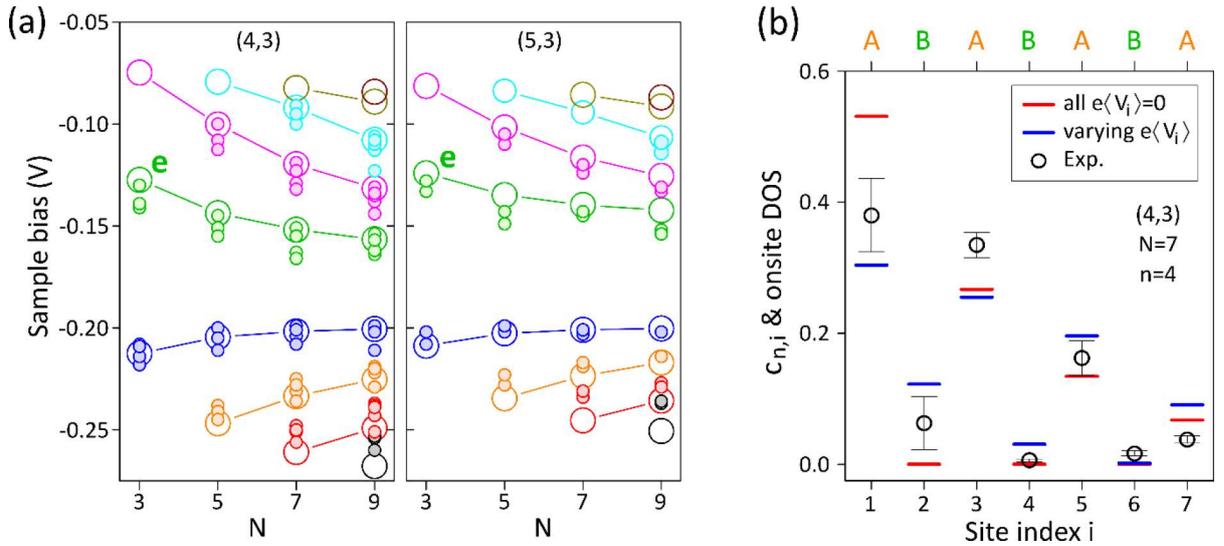

Fig. 8. (a) Experimental energies of the molecular states (small filled circles) observed for various individual chains of (4,3) [left] and (5,3) configuration (right) plotted versus N. The scatter in the experimental data points reflects the electrostatic potential disorder due to residual charged defects in the InAs(111)A substrate. Large empty circles show the theoretical energies obtained from the tight-binding Hamiltonian discussed in Appendix B. The asymmetry in the energy spectrum is a consequence of the varying onsite potential along the chain. The energies of end state e are drawn in green for both configurations. (b) Squared wave-function coefficients versus site index i calculated for the end state of a seven-dot (4,3) chain with varying onsite energies $e\langle V_i \rangle$ (blue bars) as in (a); for comparison, the red bars indicate the theoretical result for all onsite energies equal to zero as for an ideal SSH chain. Black circles show the normalized conductance measured with the tip probing the dots of a (4,3) chain with N=7, the sum of the obtained conductance values [providing a measure of the total density of states (DOS)] was normalized to 1. Error bars indicate the deviation from the experimental mean value obtained from independent measurements on four individual (4,3) chains with N=7.

The tight-binding Hamiltonian also reproduces the experimentally observed end-state delocalization. This is shown in Fig. 8(b) for the case of $N=7$ with $t_1$ and $t_2$ adjusted to the experimental hoppings corresponding to (4,3) configuration: the blue bars indicate the squared wave-function coefficients of the end state ($n=4$) calculated with the onsite energy values derived from the approach described above while the red bars show the result obtained with all onsite energies equal to zero. In the latter case, the end state resides only on the $A$ sites while it becomes delocalized and occurs also the $B$ sites at finite and varying onsite energy. The experimentally observed onsite density of states (black circles) confirms the delocalization of the end state.

## APPENDIX C: MEAN ELECTROSTATIC SHIFT

In the discussion of Fig. 6, we used a constant energy offset $E_0$ to fit the experimental $E_n$ values observed for odd-$N$ chains to the tight-binding band structure of an infinite dimer chain. $E_0$ quantifies the electrostatic shift resulting from the overall potential arising from the dots along the chain. To estimate this quantity, we take advantage of the expectation value $\langle V_i \rangle$ representing the theoretical onsite potential at site $i$: for a given molecular state $\psi_n$, we define the shift as the sum of onsite potentials weighted by the corresponding squared wave-function coefficients $c_{n,i}^2$. The mean electrostatic shift $E_0$ is the average over all molecular states, $E_0 = (e/N) \sum_{n=1}^{N} \sum_{i=1}^{N} c_{n,i}^2 \langle V_i \rangle$. The calculated $E_0$ values are in good agreement with those derived from the fitting procedure, compare the empty circles (theory) and the small filled symbols (experiment) in the right hand-side panels of Figs. 6(a) and (b).

# Supplemental Material to:

**Topological boundary states in engineered quantum-dot molecules on the InAs(111)A surface**

Van Dong Pham[1], Yi Pan[1,2], Steven C. Erwin[3], Felix von Oppen[4], Kiyoshi Kanisawa[5], and Stefan Fölsch[1,*]

[1] *Paul-Drude-Institut für Festkörperelektronik, Hausvogteiplatz 5-7, Leibniz-Institut im Forschungsverbund Berlin e. V., 10117 Berlin, Germany*

[2] *Center for Spintronics and Quantum Systems, State Key Laboratory for Mechanical Behavior of Materials, Xi'an Jiaotong University, Xi'an 710049, China*

[3] *Center for Computational Materials Science, Naval Research Laboratory, Washington, DC 20375, USA*

[4] *Dahlem Center for Complex Quantum Systems and Fachbereich Physik, Freie Universität Berlin, 14195 Berlin, Germany*

[5] *NTT Basic Research Laboratories, NTT Corporation, 3-1 Morinosato Wakamiya, Atsugi, Kanagawa, 243-0198, Japan*

**Contents:**

---

[*]foelsch@pdi-berlin.de

## S1. Dimerized chain with a light domain wall

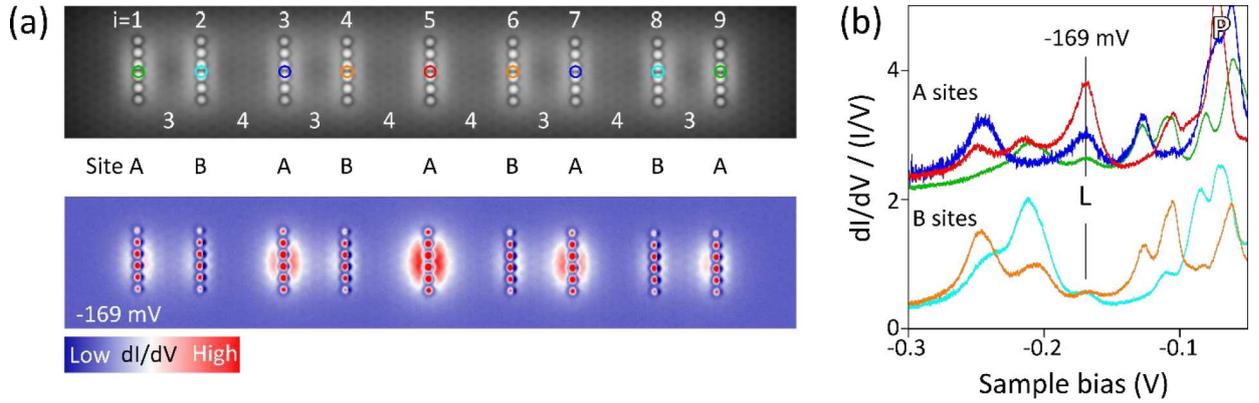

Fig. S1. (a) Upper panel: STM topography image (0.1 nA, 0.1 V) of a dimerized chain of nine dots with alternating spacing of 3 and 4 in units of √3a' =14.84 Å plus an internal domain wall at site $i$=5 belonging to sublattice $A$; the resulting chain has strongly coupled dots at both ends and a light domain wall with weak coupling between the dot at site $i$=5 and its neighbors at $i$=4 and 6. Lower panel: spatial conductance map $D(x,y)$ of the domain-wall state recorded at -169 mV (lock-in modulation voltage 10 mV peak-to-peak) revealing that the probability density is localized on the center dot at $i$=5. (b) Corresponding conductance spectra recorded with the tip probing the different dots as indicated by the colored markers in the STM image. In agreement with the $D(x,y)$ map in (a), the domain-wall state at site $i$=5 decays along the $A$ sites towards the ends of the chain while a small conductance (probability density) is detected also on the $B$ sites.